\documentclass[aps,prl,showpacs,twocolumn,groupedaddress]{revtex4-1}

\pdfoutput=1
\usepackage{amsmath}
\usepackage{cancel}
\usepackage{float}
\usepackage{ifsym}
\usepackage{color}
\usepackage{latexsym}
\usepackage{latexsym}
\usepackage{dcolumn}
\usepackage{epsfig}
\usepackage{amsfonts}
\usepackage{amsmath}
\usepackage{bm}
\usepackage{graphicx}
\usepackage{hyperref} 

\newcommand{\bnull}{{\bm{0}}}

\newcommand{\bQ}{{\bm{Q}}}

\newcommand{\bk}{{\bm{k}}}

\newcommand{\bp}{{\bm{p}}}
\newcommand{\bq}{{\bm{q}}}
\newcommand{\bu}{{\bm{u}}}

\newcommand{\bx}{{\bm{x}}}

\newcommand{\al}{\alpha}

\newcommand{\la}{\lambda}
\newcommand{\eps}{\epsilon}

\newcommand{\sig}{\sigma}

\newcommand{\La}{\Lambda}
\newcommand{\om}{\omega}

\newcommand{\calH}{{\cal H}}

\newcommand{\calF}{{\cal F}}

\begin{document}

\title{High Precision Fourier Monte Carlo Simulation of Crystalline Membranes}
\author{A.~Tr\"oster}
\email{andreas.troester@tuwien.ac.at}
\affiliation{Vienna University of Technology,
Wiedner Hauptstrasse 8-10/136,
A-1040 Wien, Austria}
\date{\today}
 
\begin{abstract}
We report an essential improvement of the plain Fourier Monte Carlo algorithm that promises to 
be a powerful tool for investigating critical behavior in a large class of lattice models, in particular 
those containing microscopic or effective long-ranged interactions.
On tuning the Monte Carlo acceptance rates separately for each wavevector, we are
able to drastically reduce critical slowing down.
We illustrate the resulting efficiency and unprecedented accuracy of our algorithm with a calculation of the universal elastic properties 
of crystalline membranes in the flat phase and derive  
a numerical estimate $\eta=0.795(10)$ for the critical exponent $\eta$ that challenges those derived from other recent simulations. 
The large system sizes accessible to our present algorithm also allow to demonstrate
that insufficiently taking into account corrections to scaling may severely hamper a finite size scaling analysis. 
This observation may also help to clarify the apparent disagreement of published numerical estimates of $\eta$ in the existing literature. 
\end{abstract}


\maketitle

Long-range interactions are ubiquitous in physics. Examples include Coulomb, dipolar and higher multipole interactions in 
traditional condensed matter physics \cite{AshcroftMermin_SSP_1976}, Wigner crystallization in fermionic quantum systems \cite{AltlandSimons2010}, 
electrostatic interactions between cold trapped ions \cite{Lewenstein2012} 
or long-range elastic interactions between
defects in condensed matter systems \cite{ChaikinLubensky_PCMP_1995}, just to name a few.
Yet, even today many aspects related to long interaction ranges are only poorly understood \cite{LongRange2008}.
Sometimes, as e.g~in many ionic systems, the long-range character is camouflaged
by screening, leaving effective short-ranged interactions (see e.g.~\cite{PatsahanMryglod_JPCM16_L235_2004}). 
However, in a number of cases one is forced to deal with the full interaction range, frequently
making theoretical attempts intractable and simulations computationally expensive.

Problems tighten further in simulations of critical long-range systems \cite{Fisher_JSP75_1_1994,Caillol_PRL_77_4093_1996}  due to the 
required large system sizes and the notorious phenomenon of critical slowing down \cite{AmitMartinMayor_FRRGCP_2005}. 
In recent years, cluster algorithms \cite{LandauBinder_MC_2009} have been designed to overcome the latter problem.
However, they may be prohibitively difficult to implement for complicated effective interactions arising e.g.~in compressible spin models \cite{BergmanHalperin_PRB13_1976,Troester_PRL100_2008}.
Yet, by utilizing the underlying translation invariance, the structure of these effective Hamiltonians often simplifies drastically on employing the Fourier transform.
Based on this observation, a radical approach was developed for lattice models in 
Refs.~\cite{Troester_PRB76_2007,TroesterDellago_F354_2007,TroesterCPC179_2008,Troester_CSSCMP_2008,Troester_PRL100_2008}
and termed Fourier Monte Carlo (FMC) algorithm.
Here we report a considerable improvement of this plain FMC algorithm, which practically also eliminates critical slowing down from the list of obstacles.
The resulting optimized Fourier Monte Carlo (OFMA) algorithm is applied to study the elastic properties of solid membranes in the flat phase with unprecedented precision.

The present article is organized as follows. We begin with a discussion the manifestations of critical slowing down 
in plain FMC and an explanation of the optimized simulation scheme suggested by this analysis.
This is followed by a short summary of the ideas underlying the description of the asymptotic elastic behavior 
of fluctuating solid membranes and the related observables
accessible in our simulations, and continued by a short explanation of our simulation setup. 
We present two approaches towards extracting a numerical estimate of the main critical exponent $\eta$ governing scaling behavior of solid
membranes at long wavelengths from the generated data:  
(i) Investigation of the correlation function of out-of-plane deformations of the membrane gives only a preliminary estimate for $\eta$ and reveals a peculiar finite size effect. 
(ii) A careful finite size scaling analysis of the membrane's mean squared displacement yields a presumably more reliable numerical result and demonstrates the importance of 
properly taking into account subleading finite size corrections.
The paper closes with a summary and short discussion of our results.      

\section{Optimized Fourier Monte Carlo}

The original idea of Fourier Monte Carlo is quite simple. In principle, any ``spin'' configuration $\{f(\bx)\}$
on a direct $d$-dim.~lattice $\Gamma$, assumed to be real for simplicity,  is in one-to-one correspondence with its set of (complex) Fourier amplitudes $\{\tilde f(\bq)\}$
defined on the (first) Brillouin zone $\tilde\Gamma$.
In FMC we completely forget about the direct lattice spins, treating the Fourier amplitudes $\tilde f(\bq)$ as our basic Monte Carlo (MC) variables. 
A MC move consists of picking a random wave vector $\bq_0\in\tilde\Gamma$ and shifting 
\begin{eqnarray}
\tilde f(\bq)\to \tilde f(\bq)+\eps\delta_{\bq,\bq_0}+\eps^*\delta_{\bq,-\bq_0},\quad |\eps|<r_\eps
\label{eqn:ncjdcnjxcnlscnckck}
\end{eqnarray}
where  $\eps$ is randomly picked from a circle of fixed radius $r_\eps$ centered around zero in the complex plane.
The tricky part is, of course, how to compute the resulting energy change $\Delta E$ accompanying the move (\ref{eqn:ncjdcnjxcnlscnckck}) in an efficient way.
Here we content ourselves with the following brief description. Harmonic terms in a lattice Hamiltionian are diagonal under Fourier transform, and therefore
it is straightforward to calculate the harmonic contribution to $\Delta E$. In contrast, an anharmonic contribution of type $\sum_{\bx}f^4(\bx)$ turns into
a sum $\sum_{\bq_1\dots\bq_4}\tilde f(\bq_1) \dots\tilde f(\bq_4)\Delta_{\Gamma}(\bq_1+\dots\bq_4)$, where the lattice delta function $\Delta_{\Gamma}(\bq)$ is defined to be $1$ if $\bq$ 
is a reciprocal lattice vector and zero else. To avoid the resulting formidable combinatorial complexity, one trivially reorganizes $\sum_\bx f^4(\bx)=\sum_{\bx}(f^2(\bx))^2$, which 
becomes diagonal in terms of the Fourier amplitudes $\widetilde{f^2}(\bq)$ of the squared field $f^2(\bx)$.
For a detailed account on how to efficiently calculate the anharmonic contribution to the energy change $\Delta E$ in terms of the amplitudes $\tilde f(\bq)$ and $\widetilde{f^2}(\bq)$  
and the nuts and bolts of FMC we refer to 
Refs.~\cite{TroesterDellago_F354_2007,TroesterCPC179_2008,Troester_CSSCMP_2008}.
 
With FMC, the number of relevant degrees of freedom can be drastically reduced in calculating universal properties
if an effective Hamiltonian defined at wave vector cutoff $\La$ is available.
Consider e.g.~a simple cubic lattice in $d$ dimensions with lattice constant $a$. Then $|q_i|\le \La_0=\pi/a$ for any $\bq\in\tilde\Gamma$.
In this case the effort of an FMC simulation at cutoff $\La<\La_0$ equals that of a 
direct lattice one with $(\La_0/\La)^d$ times more unit cells.
Furthermore, as mentioned above, translation-invariant pair interactions are diagonal under a Fourier transform and thus 
pose no problem at all, regardless of their range -- even Ewald summation is available (see Ref.~\cite{Troester_PRB_81_2_2010}).

\begin{figure}[tb]
\centering
\includegraphics[scale=0.5]{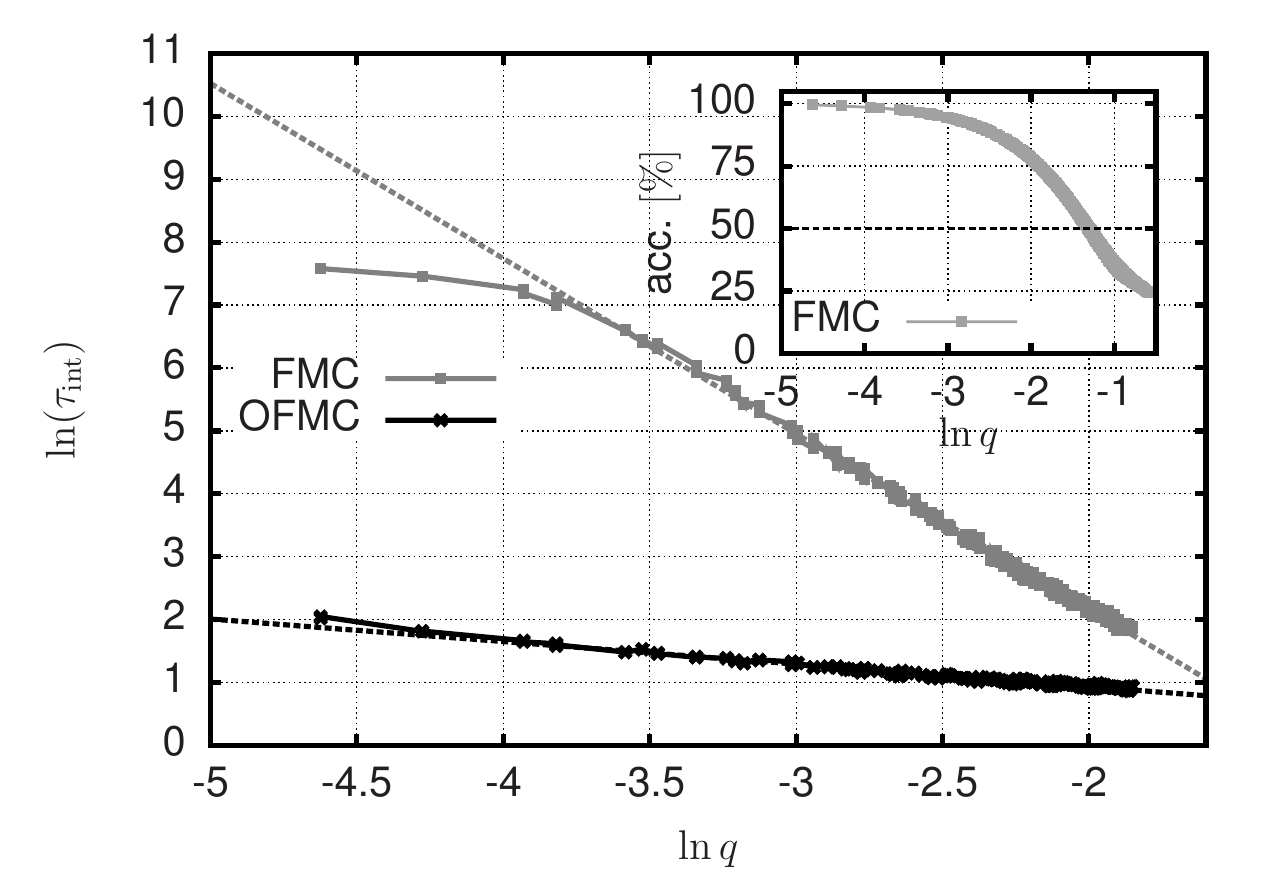}
\caption{Main plot: Dramatic reduction of integrated autocorrelation times $\tau_{\text{int}}(\bq)$ for moduli $|\tilde f(\bq)|^2$ 
for our new OFMC algorithm in comparison to plain FMC for system size $L=640$ and cutoff $\La=\pi/8$.
Deviations from linearity are due to the finite MC run time of $2^{20}$ MC steps (plain FMC)
and finite tolerance $\pm 5\%$ for a $50\%$ target acceptance (OFMC).
Inset: $q$-dependent acceptance rates as measured for plain FMC. 
}
\label{fig:tuning}
\end{figure} 

Let us now turn to to explain the announced optimization of our algorithm.
In the original version of FMC the radius $r_\eps$ for shifting the Fourier amplitudes in (\ref{eqn:ncjdcnjxcnlscnckck}) is iteratively optimized
for an average acceptance rate (AR) of, say, $30-50\%$ during the start-up of the simulation. 
Yet, even though the moves (\ref{eqn:ncjdcnjxcnlscnckck}) are collective in nature, we observe a dramatic growth of
the integrated autocorrelation times $\tau_{\text{int}}(\bq)$ \cite{BinderHeermann_2010,Berg_MCSim_2004} of the squared amplitudes $|\tilde f(\bq)|^2$ for modes
close to the critical wave vector (taken to be $\bq_c=\bnull$ for simplicity).
The sharp rise of $\tau_{\text{int}}(\bq)$ for $q=|\bq|\to\bnull$, which is the crucial quantity that determines the statistical efficiency of measuring $|\tilde f(\bq)|^2$ \cite{BinderHeermann_2010,Berg_MCSim_2004}
is the hallmark of critical slowing down. How can that be?
The average amplitudes $|\tilde f(\bq)|$ for $q\to 0$ are much larger than those for $q\gg0$, but the algorithm attempts to move them all at the same maximum pace $r_\eps$.
Thus the modes close to criticality simply make no headway in comparison to the noncritical ones, and one finds individual ARs close to $100\%$ for the relatively few ``small'' $\bq$-vectors, 
while for larger wave vectors, the numbers of which roughly increase as $\sim q^{d-1}$, ARs drop to quite low values.
However, nothing can prevent us from optimizing $r_\eps=r_\eps(\bq)$ individually for each $\bq$
in such a way that all modes $\tilde f(\bq)$ separately enjoy the same uniform AR.  In practice, since
changing $r_\eps(\bq)$ for one single $\bq$ will influence all other individual ARs in a nonlinear way,
we resort to a simple iterative procedure, aiming for a fixed collective rule-of-thumb target AR of, say, $50\%$ with a tolerance of $\pm 5\%$ during the warm-up stage of the simulation.
Excitingly, as soon as this initialization step is implemented, one observes an approximately uniform common value of $\tau_{\text{int}}(\bq)$ 
with only weak $\bq$-dependence.
Together with the collective nature of the move set (\ref{eqn:ncjdcnjxcnlscnckck}), the dramatic suppression of critical slowing down makes this ``Optimized FMC'' (OFMC) algorithm an interesting 
alternative in cases where cluster algorithms are difficult to apply.

\section{Solid Membranes}
We illustrate the benefits of the abstract strategy outlined above by considering the numerical determination of the exponent $\eta$ governing the universal elastic properties of solid membranes,
a topic of high interest in its own right in molecular biology, medicine and pharmacy, chemical synthesis, 
and soft matter physics, just to name a few scientific disciplines. Owing to the recent meteoric rise of graphene \cite{GeimNovoselov_NatMat6_2007}, 
this list has become even longer, including solid state physics, nanotechnology and electronics. 
Due to space limitations, we make no attempt to do justice to all the sophisticated theoretical and computational approaches 
that have been developed to asses membrane elasticity and merely refer to the 
authoritative references \cite{NelsonPiranWeinberg_Membranes_1988,BowickTravesset2001,Katsnelson_Graphene_2012}. Instead,
we concentrate on the fact that for a crystalline membrane, which by definition supports a nonzero static shear modulus $\mu\ne 0$,
an effective long-range interaction between its out-of-plane deformations (OPDs) emerges as follows.

Consider the so-called class of ``phantom'' membrane models for which self-avoidance effects are ignored.
Liquid phantom membranes for which $\mu=0$, are known to collapse to a rotationally invariant ``crumpled'' phase
characterized by an exponential decay of the membrane unit normal correlations \cite{DeGennesTaupin_JPC86_1982}, 
and since only short-range interactions are at work, a transition to a ``flat'' phase via spontaneous breaking of this 
continuous symmetry is ruled out by the Mermin-Wagner-Hohenberg theorem \cite{MerminWagner_PRL17_1966}.
For a crystalline membrane, however, elimination of in-plane deformations (IPDs) from the partition function by functional integration 
results in a shear-mediated effective long-range interaction
between the OPDs which Mermin-Wagner-Hohenberg has nothing to say about \cite{PelitiLeibler_PRL54_1690}. 
And indeed, at sufficiently low temperatures crystalline membranes are found to be in a ``flat'' phase,  
in which the spatial correlations of the unit normals of the membrane tend towards a nonzero constant at long distances.
In Refs.~\cite{NelsonPeliti_JPP48_1987, Nelson_Chap6_2004}, an effective Hamiltonian was formulated along these lines of thinking. 
In the so-called Monge parametrization \cite{Safran_STSIM_2003}, deformations with respect to a given two-dimensional reference plane
with coordinates $\bx=(x_1,x_2)$ are encoded in a height function $f(\bx)$ parametrizing the OPDs
and a two-dimensional vector $\bu(\bx)$ of IPDs. Variations in $f(\bx)$ give rise to a bending energy $\frac{\kappa_\La}{2}\int d^2x\left(\Delta f\right)^2(\bx)$
which is also present in liquid membranes, but for crystalline membranes $f(\bx)$ also couples to the IPD's $\bu(\bx)$ through an 
additional elastic stretching energy $\frac{1}{2}\sum_{ij}\int d^2x\left(2\mu_\La u_{ij}^2(\bx)+\la_\La u_{ii}(\bx)u_{jj}(\bx)\right)$ 
involving the Lagrangian strain tensor $u_{ij}(\bx)=[\partial_iu_j(\bx)+\partial_ju_{i}(\bx)+\partial_if(\bx)\partial_jf(\bx)]/2$.
Transforming to reciprocal space and
eliminating $\bu$ from the partition function by Gaussian integration \cite{BergmanHalperin_PRB13_1976,Troester_PRL100_2008,NelsonPeliti_JPP48_1987,Nelson_Chap6_2004},
one obtains an effective Hamiltionian
\begin{eqnarray}
\calH_\La[f]=\!\frac{\kappa_\La}{2}\int \!\frac{d^2q}{(2\pi)^2} q^4 |\tilde f(\bq)|^2
+\frac{K_\La}{8}\!\int\!\frac{d^2Q}{(2\pi)^2}|\tilde\calF(\bQ)|^2
\label{eqn:mmsksmkwswkwsmkwsmkwskwsxjsxnsjxnsjsxxm}
\end{eqnarray}
for the surviving OPD amplitudes $\tilde f(\bq\ne\bnull)$, where $K_\La=4\mu_\La(\mu_\La+\la_\La)/(2\mu_\La+\la_\La)$ is the effective 2d Young modulus at cutoff $\La$.
The amplitudes  $\tilde\calF(\bQ)$, which play a part similar to that of the amplitudes $\widetilde{f^2}(\bq)$ of the squared field $f^2(\bx)$ in the basic FMC algorithm outlined above,
are defined via the nonlocal generalized convolution in (\ref{eqn:mmsksmkwswkwsmkwsmkwskwsxjsxnsjxnsjsxxm})
\begin{eqnarray}
\tilde\calF(\bQ):=\int\frac{d^2q}{(2\pi)^2} (\hat\bQ\times\bq)^2 \tilde f(\bq)\tilde f(\bQ-\bq) 
\label{eqn:mmsksmkwswkwsmkwsmkwskwsxjsxnsjxnsjsxxm2}
\end{eqnarray}
whose wavevector dependence encodes the specific long-ranged interaction character 
(in (\ref{eqn:mmsksmkwswkwsmkwsmkwskwsxjsxnsjxnsjsxxm2}) we have formally embedded the vectors $\hat\bQ$ and $\bq$ in 3d). 
The basic observable for analyzing the thermodynamics resulting from  (\ref{eqn:mmsksmkwswkwsmkwsmkwskwsxjsxnsjxnsjsxxm}), (\ref{eqn:mmsksmkwswkwsmkwsmkwskwsxjsxnsjxnsjsxxm2})
is certainly the correlation function
of OPDs $\tilde G(\bp)\delta^{2}(\bp+\bq)\equiv\langle \tilde f(\bp)\tilde f(\bq)\rangle$.
If anharmonic contributions in (\ref{eqn:mmsksmkwswkwsmkwsmkwskwsxjsxnsjxnsjsxxm})
could be neglected, the equipartition theorem applied to the remaining harmonic bending contribution would yield
$\tilde G^{-1}(\bq)=\kappa_\La q^4$.
In reality, the anharmonicity of (\ref{eqn:mmsksmkwswkwsmkwsmkwskwsxjsxnsjxnsjsxxm}) causes the bare bending rigidity to be renormalized, and 
$\kappa_\La$ picks up a nontrivial $\bq$-dependence, such that to leading order 
\begin{eqnarray}
\tilde G^{-1}(\bq)=\kappa_\La(\bq) q^4,\qquad \kappa_\La(\bq)\sim q^{-\eta}  
\label{eqn:nnsnsxsncncncnnjnancG}  
\end{eqnarray}
The exponent $\eta$ is the central quantity governing the universal long distance elastic behavior of the flat phase.
For instance, the mean square height fluctuations $\langle(\Delta f)^2\rangle$  diverge for $L\to\infty$ like
\begin{eqnarray}
\langle(\Delta f)^2\rangle=G(\bnull)=\int\frac{d^2q}{(2\pi)^2}\tilde G(\bq)\sim L^{2\zeta}
\label{eqn:sxnnklxmkxmkxmklxlxnxxn}
\end{eqnarray}
with the roughness exponent \cite{Bowick_inNelson_2004} $\zeta=1-\eta/2$, as can be seen from
explicitly calculating the integral (\ref{eqn:sxnnklxmkxmkxmklxlxnxxn}) using an infrared cutoff $|q_i|\ge 2\pi/L$ and 
(\ref{eqn:nnsnsxsncncncnnjnancG}).

\section{Simulation Setup}
In order to numerically determine $\eta$,  
we performed OFMC simulations on a $d=2$ square lattice and we monitored the set of squared moduli $|\tilde f(\bq)|^2$ for wave vectors $\bq$ inside a suitably chosen cutoff $\La$,
together with $(\Delta f)^2\sim\sum_{\bq} |\tilde f(\bq)|^2$ and the total energy $E$ of the system. 
From the corresponding raw data series of these observables, which we also analyzed to ensure the complete equilibration of our simulations before MC measurements, 
we calculated estimates of the corresponding statistical errors and the resulting integrated autocorrelation times
$\tau_{\text{int}}(\bq)$, $\tau_{(\Delta f)^2}$ and $\tau_E$ using the jackknife approach \cite{Berg_MCSim_2004,AmitMartinMayor_FRRGCP_2005}.
These estimates were cross-checked for consistency by directly determining 
$\tau_{\text{int}}(\bq)$, $\tau_{(\Delta f)^2}$ and $\tau_E$ from the raw data autocorrelation functions
\cite{BinderHeermann_2010,Berg_MCSim_2004} and double-cross-checked
using the blocking method \cite{FlyvbjergPetersen_JCP91_461_1989,AmitMartinMayor_FRRGCP_2005}.
The choice of renormalized parameters $\kappa_\La=0.1,\, K_\La=1.0$, in which a factor $(k_BT)^{-1}$ has been absorbed, 
was motivated by the heuristic principle to have an approximate balance of harmonic and anharmonic contributions 
to total average energy changes.
Compared to plain FMC, one indeed observes the expected tremendous reduction of 
autocorrelation times $\tau_{\text{int}}(\bq)$ 
(cf.~Fig.~\ref{fig:tuning}), $\tau_{(\Delta f)^2}$ and $\tau_E$ (not shown).
After completing this work we realized that an optimization similar to ours had also been attempted in Ref.~\cite{LosFasolino_PRB80_121405_2009}, 
but was only implemented in the trivial case of a simple quasiharmonic model, and their collective ``wave vector moves'' were carried out at a frequency of 
one MC sweep on average, merely complementing a localized real-space MC move set. Suppression of critical slowing down with an efficiency comparable to that of our present method was therefore clearly out of reach.
\begin{figure}[tb]
\centering
\includegraphics[scale=0.5]{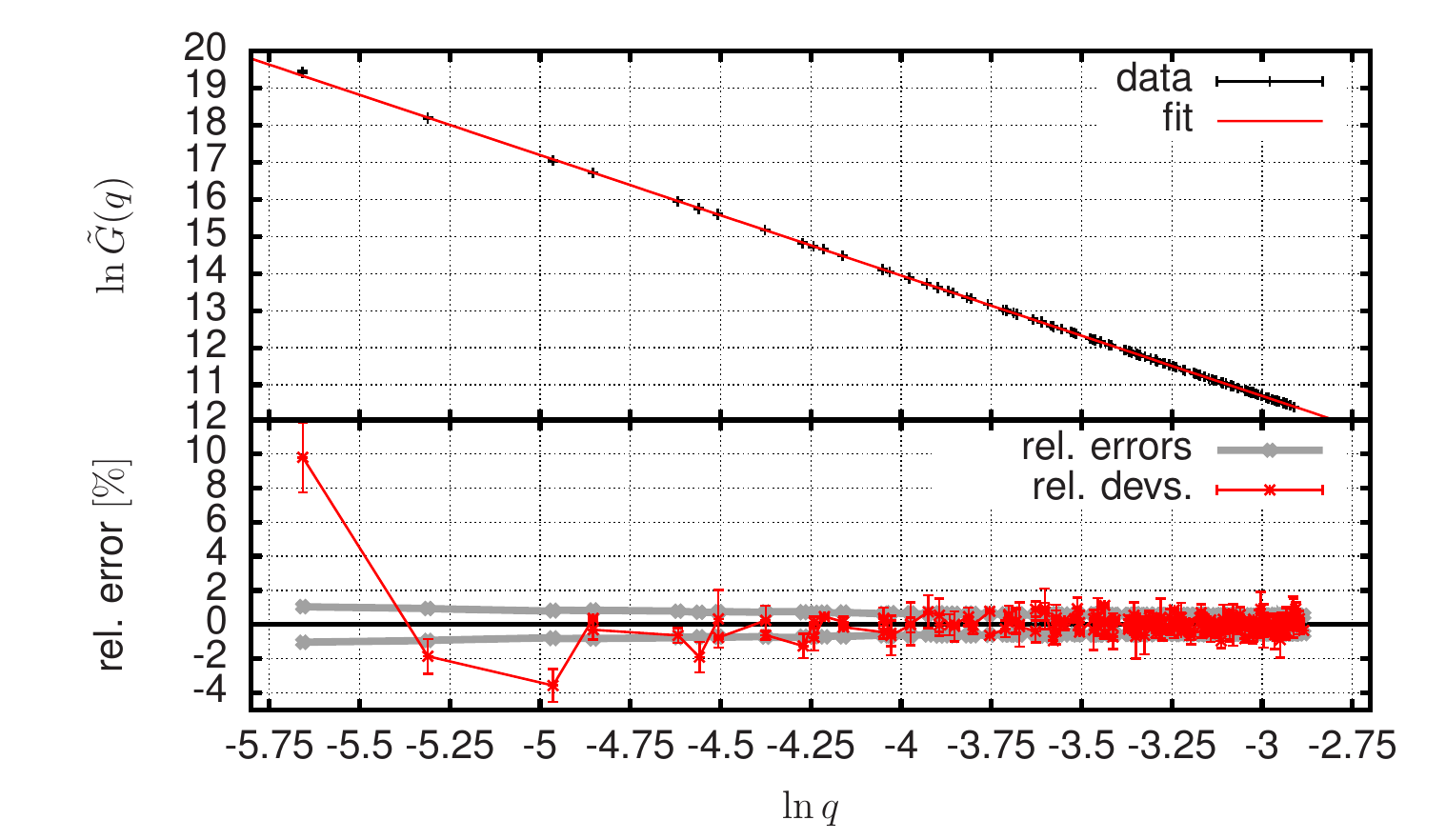}
\caption{Top: fit of simulation results for $\tilde G(\bq)$
at linear system size $L=1800$ and cutoff $\La=\pi/15$. Error bars are smaller than symbol size.
Bottom: Relative deviations of simulation data from the fit. Red lines are a guide to the eye.
Gray lines indicate the relative statistical errors of the simulation data in percent.
}
\label{fig:loglogGvonkFull}
\end{figure} 

\section{Out-Of-Plane Deformation Correlations}

\begin{figure}[tb]
\hspace{-1.9cm}\includegraphics[clip=true,scale=0.5]{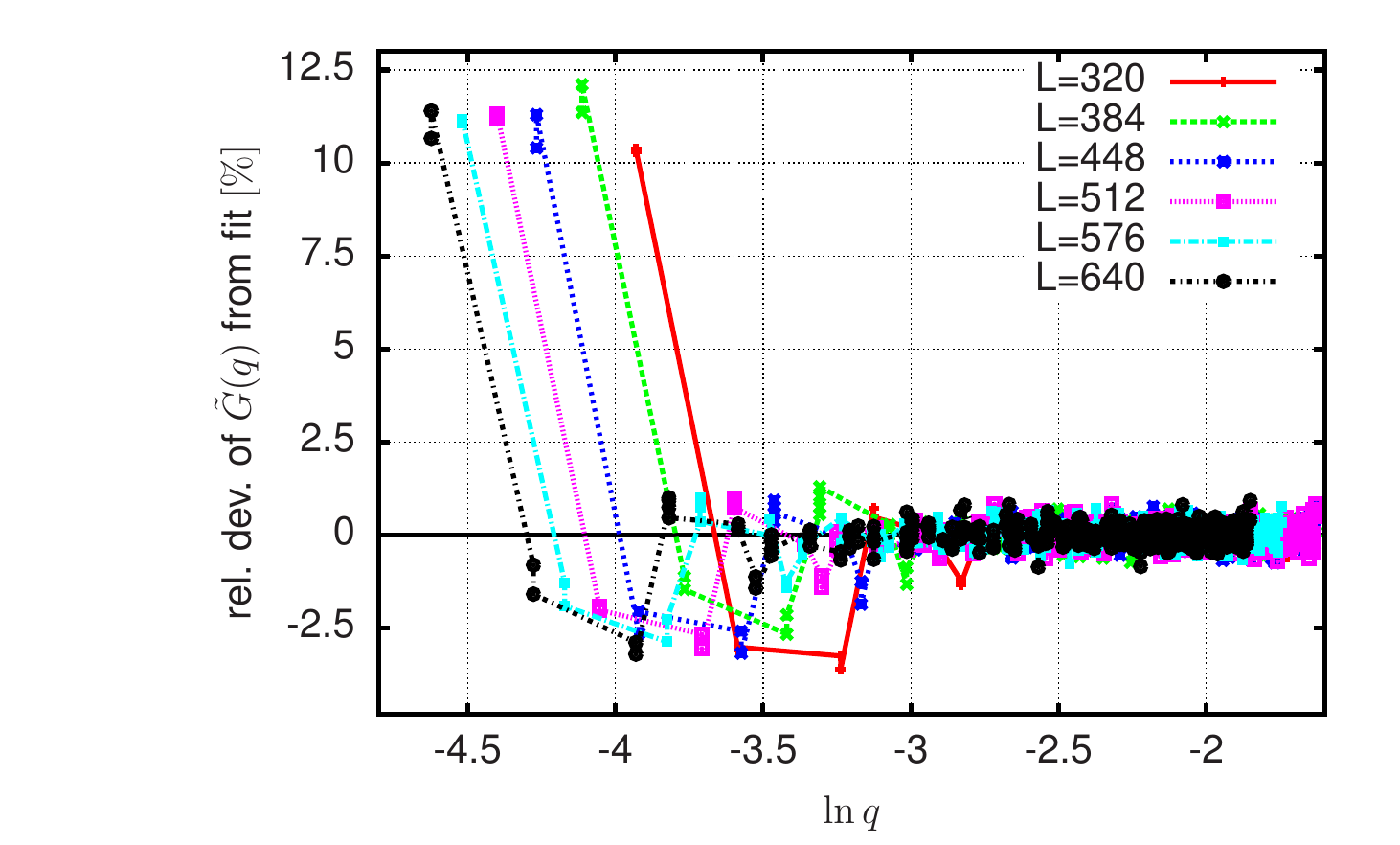}
\caption{Comparison of relative deviations of simulation data for $\tilde G(\bq)$   
for various values of $L$ and cubic cutoff $\La=\pi/8$.
}
\label{fig:GvonkreldevsCollection}
\end{figure}

\begin{figure}[tb]
\centering
\includegraphics[scale=0.55]{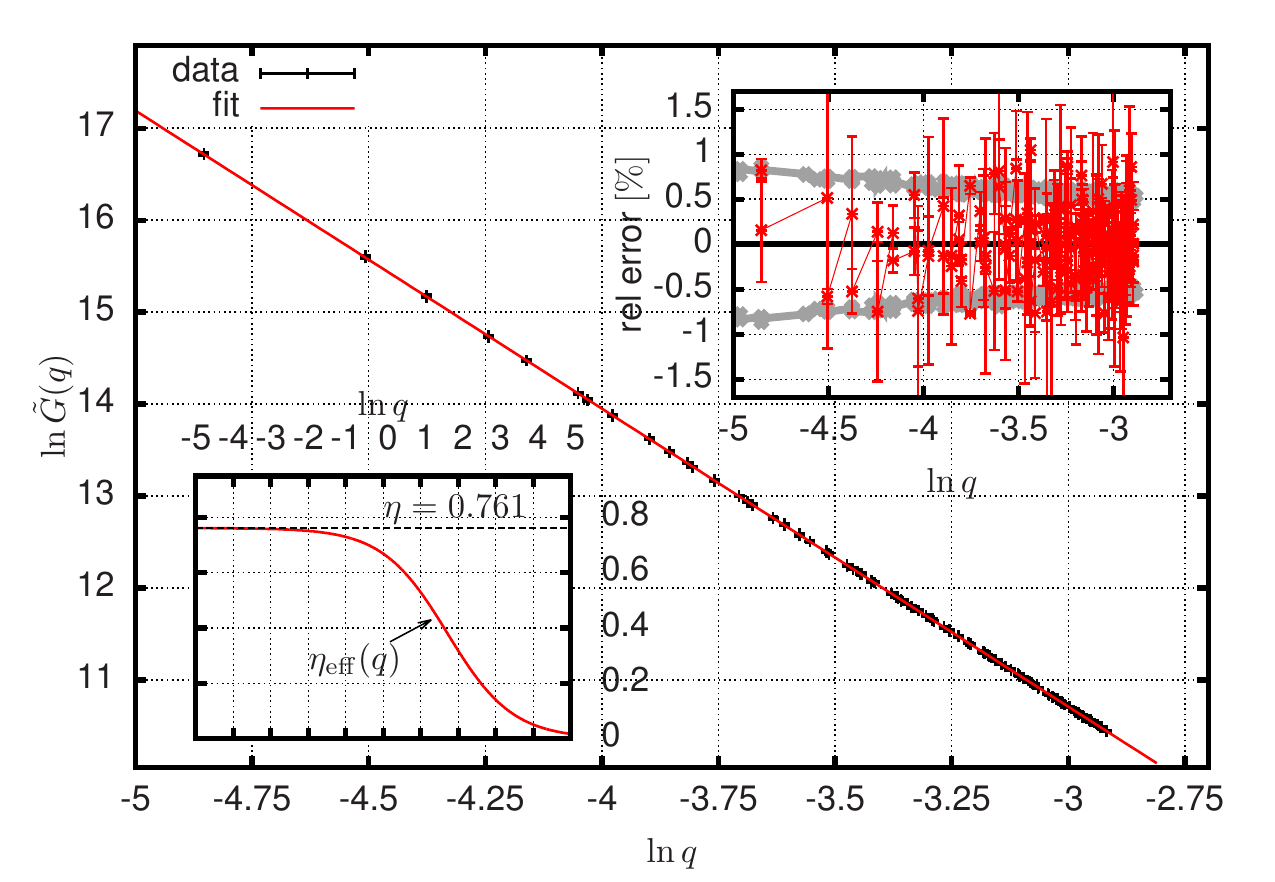}
\caption{Main plot: Same as top of Fig.~\ref{fig:loglogGvonkFull}, but
 with all contributions for $\bq$-vectors of symmetry type $(1,0)$ and $(1,1)$ omitted. 
Right upper inset: relative deviations of simulation data from the fit. Red lines are a guide to the eye.
Gray lines indicate the relative statistical errors of the simulation data in percent.
Left lower inset: crossover function (\ref{eqn:sxxmsmxcssoslcc}).}
\label{fig:loglogGvonkSelection}
\end{figure}

Unfortunately, trying to numerically extract $\eta$ from the correlation function $\tilde G(\bq)$ of the OOP deformations via (\ref{eqn:nnsnsxsncncncnnjnancG}) with sufficient precision suffers from several caveats.
First of all, for $\tilde G(\bq)$ no systematic FSS machinery seems to be available. Thus, even for a very large system size, simulation results may be contaminated by a 
residual finite size dependence and effects of lattice anisotropy.
Also, the quality of results may hinge crucially on the particular functional form of ansatz for the 
effective exponent function required for interpolating between the mean field (MF) value $\eta=0$ and the critical value of $\eta$ (cf~e.g.~Ref.~\cite{LuijtenBinder_PRE58_R4060_1998}).
Integration of our ansatz 
\begin{eqnarray}
\eta_{\text{eff}}(q)\equiv\frac{\eta}{1+\al q^{\sig}}
\label{eqn:sxxmsmxcssoslcc}
\end{eqnarray}
with two parameters $\sig$ and $\al$ besides $\eta$ to allow adjusting both position and width of the crossover,
yields the fit function $\tilde G^{-1}(\bq) =\frac{\kappa_\La q^4}{k_BT}\cdot \left[1+(\al q^{\sig})^{-1}\right]^{\eta/\sig}$.
Derived from a parallel version of our FMC code which will be described elsewhere, 
Fig.~\ref{fig:loglogGvonkFull} shows the typical results of a corresponding fit obtained for a large system of linear size $L=1800$ and cutoff $\La=\pi/15$,
effectively resembling a direct lattice system of linear size $L_0=1800\times15=27000$.
At first glance, the fit looks quite acceptable, producing a value of
$\eta=0.751(17)$. However, comparing the relative statistical errors of the individual data to their relative deviations from the fit,
our high statistical accuracy allows to resolve that
most of the deviations remain within the range of the statistical noise of the data except for the smallest nonzero $\bq$-vectors
$(1,0)\cdot 2\pi/L$ and $(0,1)\cdot  2\pi/L$ on the cubic lattice, which exceed $10\%$. Following next in size are the vectors
$(1,1)\cdot  2\pi/L$ and $(1,-1)\cdot  2\pi/L$, who also show a noticeable deviation, albeit much smaller and opposite 
in tendency. To study the finite size dependence of these deviations, we decided to
investigate a collection of systems with sizes $L=32n,\ n=1,2,\dots,20,22,\dots,28$ and cubic cutoff $\La=\pi/8$.
Indeed, for growing system size $L$, the observed irregularities are qualitatively completely similar but are shifted systematically in parallel towards 
$\lim_{L\to\infty}2\pi/L=\bnull$ (see Fig.~\ref{fig:GvonkreldevsCollection}). Having ruled out trivial explanations for the observed behavior by various consistency checks, we conclude
that what see is a finite size effect related to the anisotropic structure of the convolution (\ref{eqn:mmsksmkwswkwsmkwsmkwskwsxjsxnsjxnsjsxxm2}).
To extract a numerical estimate of $\eta$ from $\tilde G(\bq)$, we thus decided to discard 
most deviatoric data, i.e.~those for $\bq\parallel (1,0),(0,1),(1,1),(1,-1)$.
As Fig.~\ref{fig:loglogGvonkSelection} shows, the resulting fit is excellent and produces 
$\eta=0.761(8)$. In view of the encountered difficulties, however, this estimate may be contaminated by residual systematic errors beyond the pure statistical error of the fit we report here, as 
we shall argue below. Nevertheless, to appreciate the quality of our present data, note  that the crossover (\ref{eqn:sxxmsmxcssoslcc}) of $\tilde G(\bq)$ from MF to critical behavior roughly  
occurs at a ``Ginzburg wave vector'' \cite{Katsnelson_Graphene_2012} $q_G\approx\sqrt{3k_BTK_\La/8\pi\kappa_\La^2}$
analytically defined by the breakdown of the harmonic approximation. The ability of a simulation to efficiently sample the scaling region thus depends on the ratio $\rho=q_G/(2\pi/L)$ of $q_G$ 
to the smallest accessible wave vector component $2\pi/L$.
For instance, in the case of the atomistic MC simulations presented in Ref.~\cite{LosFasolino_PRB80_121405_2009} $\rho$ turns out to be only 
approximately $9$. In contrast, for our largest systems we obtain $\rho=q_G/(2\pi/L)\approx 66$, thus providing $\rho\approx (66/9)^2\approx 54$ times more data that actually explore the scaling region.

\section{Out-Of-Plane Mean Squared Deformations}
In contrast to $\tilde G(\bq)$, analysis of $\langle(\Delta f)^2\rangle$ allows to employ finite size scaling (FSS) techniques. 
On our discrete lattice, the integral in (\ref{eqn:sxnnklxmkxmkxmklxlxnxxn}) is replaced by a sum, whose asymptotic scaling behavior should comply to the
general form (\ref{eqn:sxnnklxmkxmkxmklxlxnxxn}). By definition, an asymptotic scaling law of type (\ref{eqn:sxnnklxmkxmkxmklxlxnxxn})
allows for various subleading algebraic and logarithmic corrections at finite $L$, we should therefore be included 
in a fit to the data in order to obtain precise estimates for both $\eta$ and the corresponding error $\sigma_\eta$ (see e.g.~\cite{AmitMartinMayor_FRRGCP_2005}). 
Unfortunately, however, to the author's best knowledge the structure of these corrections has not 
been worked out analytically up to date. We therefore have to allow for a priori unknown subleading corrections of logarithmic as well as power law type.
In addition, analytical and numerical tests based on our crossover ansatz (\ref{eqn:sxxmsmxcssoslcc}) suggest the inclusion of a constant $\delta>0$,
such that we arrive at a FSS ansatz for $(\Delta f)^2\sim\sum_{\bq} |\tilde f(\bq)|^2$ of type
\begin{eqnarray}
(\Delta f)^2\sim \delta+\alpha L^{2-\eta}(1+\beta \ln L+\gamma L^{-\om}) 
\label{eqn:bchbqnqjnxqjxnqsnbqdqwi}
\end{eqnarray}
in which we limit ourselves to including a single positive algebraic correction with exponent $\om$. However, since the trade-off between logarithmic and algebraic 
corrections for small $\om$ makes it numerically difficult to obtain meaningful fits, we content ourselves to studying both corrections separately.

\begin{figure}[tb]
\centering
\includegraphics[scale=0.5]{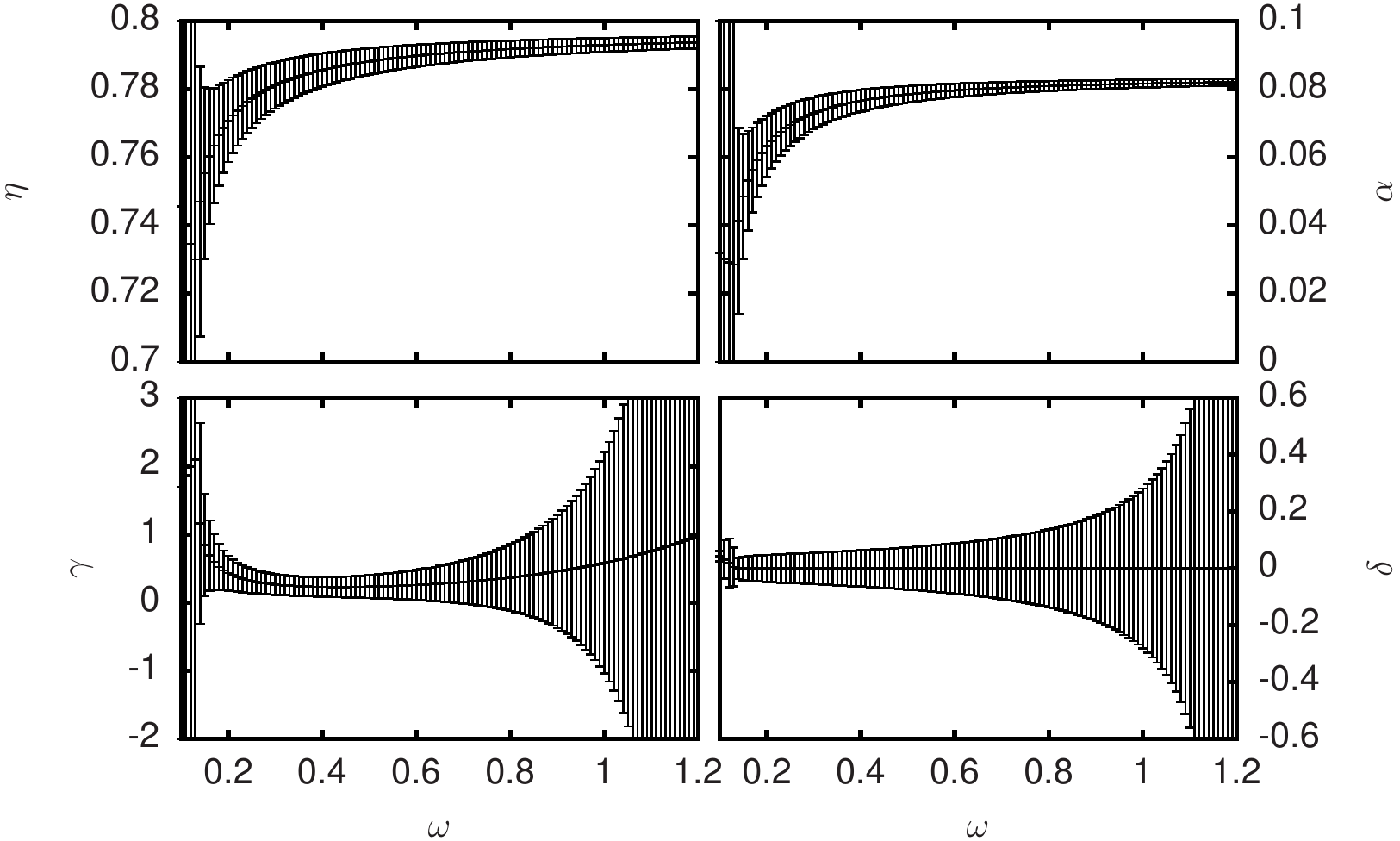}
\caption{Estimates and accompanying statistical errors obtained for the target exponent $\eta$ and the fit parameters $\al,\gamma,\delta$ 
as obtained from fits of the FSS ansatz (\ref{eqn:bchbqnqjnxqjxnqsnbqdqwi}) for $\beta=0$ at various values of $\om\in[0.1,1.2]$. 
\label{fig:OmegaDependence}}
\end{figure}

\begin{figure}[tb]
\centering
\includegraphics[scale=0.5]{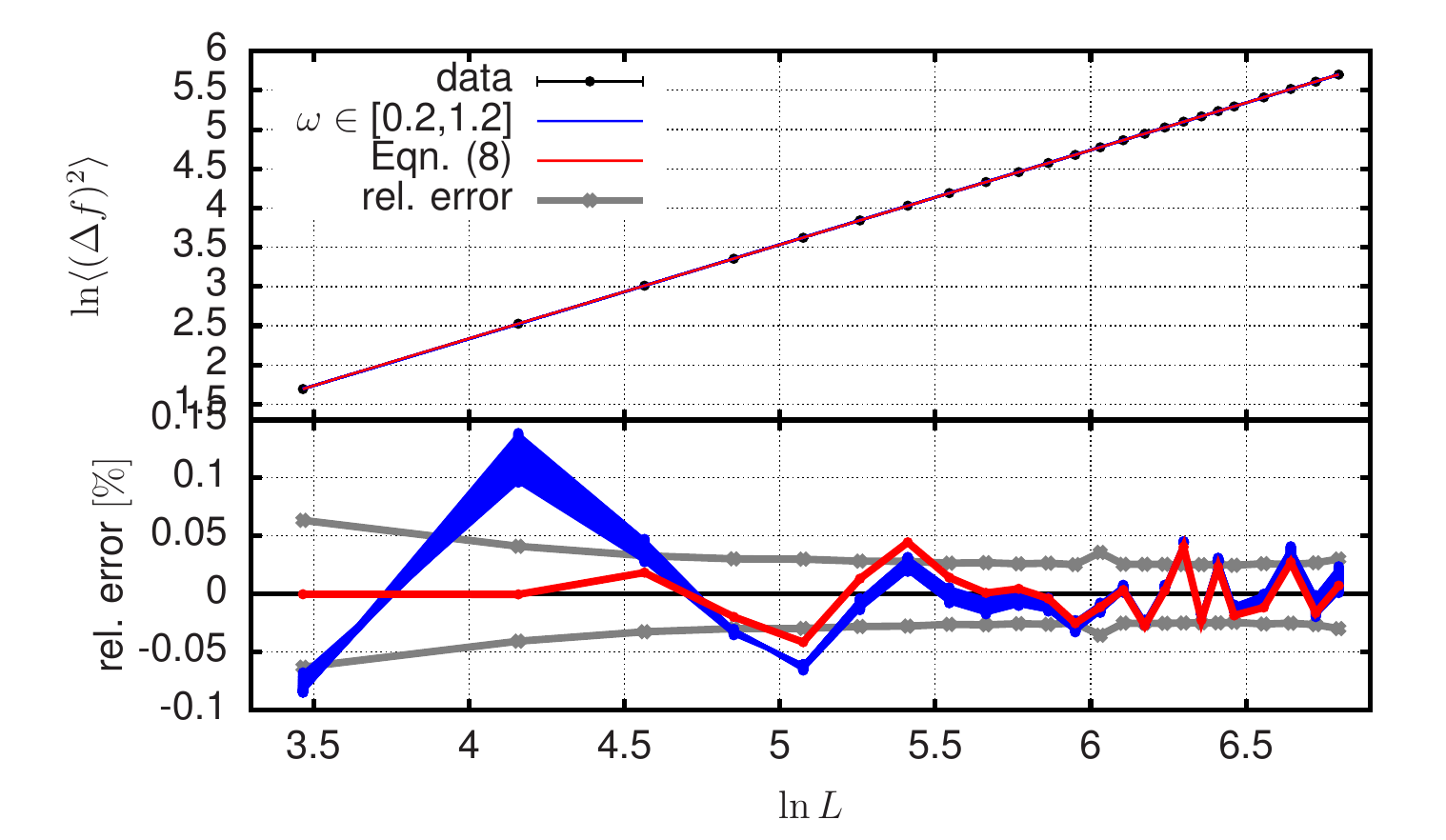}
\caption{Top: fits of Eqs.~(\ref{eqn:bchbqnqjnxqjxnqsnbqdqwi}) for various fixed values $om$ taken from the interval
$[0.1,1.2]$ and Eqn.~(\ref{eqn:bchbqnqjnxqjxnqsnbqdqwinaiv}) 
to simulation results for $\langle(\Delta f)^2\rangle$
obtained for $L=32n,\ n=1,2,\dots,20,22,\dots,28$ and $\La=\pi/8$. Data error bars are smaller than symbol size. All fits are practically indistinguishable at this scale.
Bottom: Relative deviations of simulation data from the fits. Lines are a guide to the eye.
Gray lines indicate the relative statistical errors of the simulation data in percent.
}
\label{fig:etafromDeltafquad}
\end{figure}

A purely logarithmic correction yields the value $\eta=0.793$, but unfortunately this result does not inspire much confidence since the corresponding statistical fitting error is of the order of
$10^5$. Turning to algebraic corrections, we fix $b=0$ and attempt to fit (\ref{eqn:bchbqnqjnxqjxnqsnbqdqwi}) to the data with a variable correction exponent $\om$, which yields
the vague result $\eta\approx 0.781(100)$ for $\eta$, accompanied by the much too imprecise estimate $\om=0.372\pm 2.4$ for $\om$. 
A fit based on fixing $\om$ to this value produces $\eta=0.784(5)$, but the dependence on the choice of $\omega$ remains unclear. To investigate this problem further,
we thus decided to fit the data using the ansatz (\ref{eqn:bchbqnqjnxqjxnqsnbqdqwi}) for a range of values of $\om$.
We observe that only for roughly $\om\in[0.2,1.2]$ these fits produced meaningful values for the parameters $\alpha,\gamma$ and $\delta$ within equally meaningful uncertainties.    
The results, which are gathered in Fig.~\ref{fig:OmegaDependence}, illustrate the unpleasant fact that the indeterminacy of $\om$ not only affects the quality of the error estimates for
the sought-after exponent $\eta$ but indeed has a non-negligible effect on the estimated value of $\eta$ itself. Indeed, Fig.~\ref{fig:OmegaDependence} makes it obvious that 
without precise knowledge of the exponent $\om$, FSS based on the ansatz (\ref{eqn:bchbqnqjnxqjxnqsnbqdqwi}) does not allow to extract a reliable result for $\eta$.
Moreover, the lower part of Fig.~(\ref{fig:etafromDeltafquad}) demonstrates that, regardless of which value of $\om\in[0.2,1.2]$ is chosen, residual deviations between fits 
based on Eqn.~(\ref{eqn:bchbqnqjnxqjxnqsnbqdqwi}) and the actual data at smaller $L$ are found to persist. In combination with the pronounced $\om$-dependence of the obtained values for $\eta$, 
this provides compelling evidence that an ansatz of type (\ref{eqn:bchbqnqjnxqjxnqsnbqdqwi}) does not properly account for the finite size corrections to scaling.

In contrast, an alternative fit based on the somewhat simple-minded ad-hoc ansatz
\begin{eqnarray}
(\Delta f)^2\sim\delta+\alpha L^{2-\eta}(1+\beta/L+\gamma/L^{2}) 
\label{eqn:bchbqnqjnxqjxnqsnbqdqwinaiv}
\end{eqnarray}
not only yields a comparably good agreement at large $L$ (where any reasonable ansatz for $(\Delta f)^2$ with the correct built-in asymptotics will works equally well)
but apparently also captures the behavior at small $L$ with an accuracy that seems very hard to improve any further
(cf.~again the lower part of Fig.~(\ref{fig:etafromDeltafquad})). With this surprisingly good numerical agreement at both large and small $L$, 
it therefore appears to be of little relevance that the assumed representation of the 
scaling corrections as mere inverse integer powers of $L$ lacks a strict theoretical justification. Numerically, a fit using (\ref{eqn:bchbqnqjnxqjxnqsnbqdqwinaiv}) produces the value
$\eta=0.795(5)$. Including a conservative safety margin of a factor of two in the error bar to this result, we are finally led to report the fair estimate
\begin{eqnarray}
\eta = 0.795(10)
\end{eqnarray}


\begin{table}[t]
\caption{Selection of numerical results for $\eta$. Question marks indicate unreported error bars or possible systematic errors.}
\begin{center}
\begin{tabular}{clc}
\hline
$\eta$ & Method & Ref.\\
\hline
$0.750(5)$   &MC, Gaussian spring pot., IPDs     &\cite{Bowick_JPF6_1321_1996}\\
$0.72(4)$    &MC, Gaussian spring pot., OPDs     &\cite{Bowick_JPF6_1321_1996}\\
$0.849(?)$   &nonperturbative RG                 &\cite{KownackiMouhanna_PRE79_040101_2009}\\
$0.821(?)$   &self-consistent field approx.      &\cite{LeDoussalRadzihovsky_PRL69_1992}\\
$0.85(?)$    &MC, atomistic carbon potential     &\cite{FasolinoLosKatsnelson_NM6_2007}\\
$0.85$(?)    &MC, MD, quasiharmonic model        &\cite{LosFasolino_PRB80_121405_2009}\\
$0.761(?)$   &OFMC, $\tilde G(\bq)$              &this work\\
$0.795(10)$  &OFMC, $\langle(\Delta f)^2\rangle$ &this work\\
\hline 
\end{tabular}
\end{center}
\label{tab:etaliterature}
\end{table}

\section{Summary and Discussion}

In summary, we have explained how to optimize the plain Fourier MC algorithm and effectively eliminate critical slowing down.
As an application, we obtain high precision simulation results for the universal elastic behavior of
a crystalline membrane in the flat phase. For this problem, a considerable dispersion of previous estimates 
of $\eta$ has been published in the existing literature (cf. Table \ref{tab:etaliterature}) 
for comparison with our present estimates), and even a complete violation of scaling has been claimed \cite{FasolinoLosKatsnelson_NM6_2007}.
While our present simulations fully support the conventional universal scaling theory of solid membranes,
our high precision and the accessibility of unprecedented effective system sizes allow to gain some new insight into the reason for the remaining numerical 
discrepancies in the published results for $\eta$. Indeed, we find that the analysis of both $\tilde G(\bk)$ and $\langle(\Delta f)^2\rangle$ 
requires an extremely careful analysis of finite size corrections. In the case of $\tilde G(\bk)$, finite size effects 
are serious enough to cast any attempt to extract $\eta$ from a naive scaling fit into doubt, but 
a finite size scaling analysis of $\langle(\Delta f)^2\rangle$ 
may also be severely biased by ignoring or incompletely taking into account subleading corrections. Our analysis reveals that such corrections clearly do affect 
the numerics of our present work even though the effective system sizes that we are able to access are much larger than those of previous studies. 
We believe that these finding may also serve to explain the mentioned disagreement on the published numerical values for $\eta$ in the existing literature that is immediately apparent from
a glance at Table \ref{tab:etaliterature}. 
 
The presented strategy for suppressing critical slowing is expected to work equally well for a large class of 
other lattice models at or near criticality. As to elastic membranes, the specific approach of the present paper should also be equally applicable to the hexatic case, for which 
we hope to obtain first results in the near future.

\begin{acknowledgments}
We would like to thank K.~Binder, C.~Dellago, W.~Janke, G.~Kahl, M.~Katsnelson, U.~Pedersen and A.~Travesset for discussions and
acknowledge support by the Austrian Science Fund (FWF) Project P22087-N16.
Major parts of our computations were performed on the Vienna Scientific Cluster (VSC2). 
\end{acknowledgments}



\end{document}